\documentclass[twocolumn,showpacs,preprintnumbers,amsmath,amssymb,superscriptaddress]{revtex4}%

\usepackage{graphicx}
\usepackage{dcolumn}
\usepackage{bm}
\usepackage{color}
\usepackage[version=3]{mhchem}

\begin{document}

\preprint{preprint(\today)}
\title{Linear Scaling of the Superfluid Density with the Critical Temperature in the Layered Superconductor 2H-NbSe$_{2}$}

\author{F.O.~von~Rohr}
\affiliation{Department of Chemistry, University of Z\"{u}rich, CH-8057 Z\"{u}rich, Switzerland}
\affiliation{Physik-Institut der Universit\"{a}t Z\"{u}rich, Winterthurerstrasse 190, CH-8057 Z\"{u}rich, Switzerland}

\author{J.-C.~Orain}
\affiliation{Laboratory for Muon Spin Spectroscopy, Paul Scherrer Institute, CH-5232 Villigen PSI, Switzerland}

\author{R.~Khasanov}
\affiliation{Laboratory for Muon Spin Spectroscopy, Paul Scherrer Institute, CH-5232 Villigen PSI, Switzerland}

\author{Z.~Shermadini}
\affiliation{Laboratory for Muon Spin Spectroscopy, Paul Scherrer Institute, CH-5232 Villigen PSI, Switzerland}

\author{A.~Nikitin}
\affiliation{Laboratory for Muon Spin Spectroscopy, Paul Scherrer Institute, CH-5232 Villigen PSI, Switzerland}

\author{J.~Chang}
\affiliation{Physik-Institut der Universit\"{a}t Z\"{u}rich, Winterthurerstrasse 190, CH-8057 Z\"{u}rich, Switzerland}

\author{A.R.~Wieteska}
\affiliation{Department of Physics, Columbia University, New York, NY 10027, USA}

\author{A.N.~Pasupathy}
\affiliation{Department of Physics, Columbia University, New York, NY 10027, USA}

\author{M.Z.~Hasan}
\affiliation{Laboratory for Topological Quantum Matter and Spectroscopy, Department of Physics, Princeton University, Princeton, New Jersey 08544, USA}

\author{A.~Amato}
\affiliation{Laboratory for Muon Spin Spectroscopy, Paul Scherrer Institute, CH-5232 Villigen PSI, Switzerland}

\author{H.~Luetkens}
\affiliation{Laboratory for Muon Spin Spectroscopy, Paul Scherrer Institute, CH-5232 Villigen PSI, Switzerland}

\author{Y.J.~Uemura}
\affiliation{Department of Physics, Columbia University, New York, NY 10027, USA}

\author{Z.~Guguchia}
\email{zurab.guguchia@psi.ch}\affiliation{Laboratory for Muon Spin Spectroscopy, Paul Scherrer Institute, CH-5232 Villigen PSI, Switzerland}
\affiliation{Department of Physics, Columbia University, New York, NY 10027, USA}
\affiliation{Laboratory for Topological Quantum Matter and Spectroscopy, Department of Physics, Princeton University, Princeton, New Jersey 08544, USA}

\begin{abstract}
We report on high-pressure ($p_{\rm max}$ = 2.1 GPa) muon spin rotation experiments probing the temperature-dependent magnetic penetration depth $\lambda\left(T\right)$ in the layered superconductor 2H-NbSe$_{2}$. Upon increasing the pressure, we observe a substantial increase of the superfluid density $n_{s}/m^{*}$ ${\propto}$ 1/${\lambda}^{2}$, which we find to scale linearly with $T_{c}$. This linear scaling is considered a hallmark feature of unconventional superconductivity, especially in high-temperature cuprate superconductors. Our current results, along with our earlier findings on 1T'-\ce{MoTe2} (Z. Guguchia et. al., Nature Communications 8, 1082 (2017)), demonstrate that this linear relation is also an intrinsic property of the superconductivity in transition metal dichalcogenides, whereas the ratio $T_{\rm c}$/$T_{\rm F}$ is approximately a factor of 20 lower than the ratio observed in hole-doped cuprates. We, furthermore, find that the values of the superconducting gaps are insensitive to the suppression of the quasi-two-dimensional CDW state, indicating that the CDW ordering and the superconductivity in 2H-NbSe$_{2}$ are independent of each other.   
\end{abstract}
\maketitle


Transition metal dichalcogenides (TMDs) are a class of layered materials, which presently attract great interest because of their versatile physical, chemical, and mechanical properties\cite{Soluyanov,Kaminski,Xu,Ali1,QiCava,Qian,Zhang,Ugeda,Bozin,SunY,Hasan1}. The TMDs share the \ce{MX2} formula, with M being a transition metal (M = Ti, Zr, Hf, V, Nb, Ta, Mo, W or Re) and X being a chalcogen (X = S, Se, or Te). These compounds crystallize in different structural phases resulting from different stacking of the individual \ce{MX2} layers, with van der Waals bonding between them. \\

A variety of novel physical phenomena have been reported in TMDs systems. Most recently topological physics with Dirac-type dispersion, exotic optical and transport behavior originating from valley splitting were predicted and observed \cite{Soluyanov,Ali1}. Correlated physics has been studied intensively especially in 2H-\ce{TaS2}, 2H-\ce{NbSe2}, and more recently in 1T'-\ce{MoTe2} \cite{Sipos,Cho,Kanatz,Luo17,QiCava}. 2H-\ce{NbSe2} is a superconductor with a critical temperature $T_{\rm c} \approx$ 7.3~K, and hosts a two-dimensional charge density wave (CDW) with a critical temperature of $T_{\rm CDW} \approx$ 33 K \cite{Morris,Moncton,Du,Berthier,Littlewood1981,Littlewood1982}. The superconducting transition temperature of 2H-\ce{NbSe2} is remarkably high in comparison with other TMDs, where commonly transition temperatures below 4~K are observed. Recently, it was found that both the superconducting state, as well as the CDW ordering remain intact even for a single layer of this material \cite{Ugeda}. The superconducting transition temperature is thereby lowered in the 2D limit to $T_{\rm c} \approx$ 1.1~K. Previously, the effect of hydrostatic pressure on the CDW ordering and superconductivity were studied in 2H-NbSe$_2$ by means of magnetization and resistivity measurements \cite{YFeng,MLeroux}. It was found that upon increasing pressures, CDW ordering is suppressed. The suppression of charge-density wave ordering commonly causes a strong increase of the superconducting transition temperature $T_{\rm c}$ (see, e.g. references \cite{vonrohr1,vonrohr2,Nozaki}). The generic occurrence of a ``fragile`` SC phase in the systems with competing SC and CDW ground states was reported, recently \cite{KivelsonCDW}. In 2H-\ce{NbSe2}, $T_{\rm c}$ increases only slightly ($\Delta T_{\rm c} {\textless}$ 1 K) with increasing pressure without any anomaly across the critical pressure at which the CDW state disappears. Moreover, both CDW order parameter amplitude as well as its static phase coherence length are unaffected when the superconductivity is suppressed in a high magnetic field \cite{Du}. Recent X-ray experiments also suggest that different phonon branches are responsible for SC and CDW orders \cite{MLeroux}. These findings suggested that superconductivity is only weakly affected by the suppression of the CDW ordering 
   \cite{Du,MLeroux,Berthier,Littlewood1981,Littlewood1982}.\\

\begin{figure*}[t!]
\centering
\includegraphics[width=1.3\linewidth]{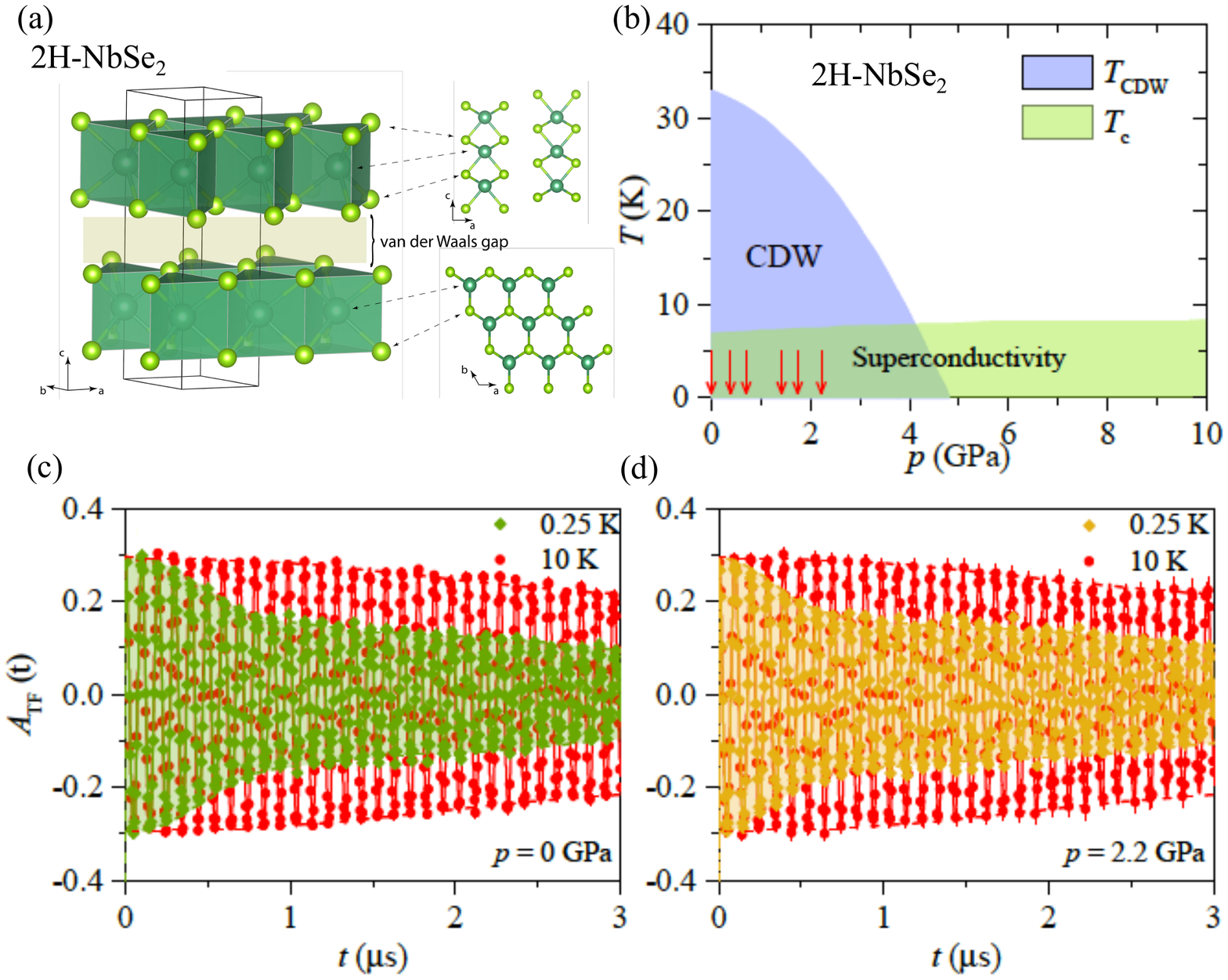}
\vspace{-2.0cm}
\caption{\textbf{Schematic Phase Diagram and Transverse-field (TF) ${\mu}$SR time spectra for NbSe$_{2}$.} 
(a) Pressure dependence of $T_{\rm c}$ and the CDW temperature $T_{\rm CDW}$ (ref. \cite{QiCava}). The red arrows mark the pressures at which the $T$-dependence of the penetration depth was measured. The TF spectra, above and below $T_{\rm c}$ at pressures of (b) $p$ = 0 GPa and (c) $p$ = 2.2 GPa are shown. The solid lines in panel (b) and (c) represent fits to the data by means of Eq.~1; The dashed lines are guides to the eyes.}
\label{fig1}
\end{figure*}

Layered superconductors, with highly anisotropic electronic properties have been found to be potential hosts for unconventional superconductivity. The most prominent example are the cuprates, consisting of \ce{CuO2} layers and the iron pnictides superconductors, with tetrahedra coordinated \ce{Fe\textit{Pn}4} layers. Other examples for unconventional superconductivity in layered systems are a monolayer of FeSe on STO \cite{Wang} and the reported superconductivity in twisted bilayer graphene \cite{YCao}. Recently, we have reported on muon spin rotation (${\mu}$SR) measurements on the TMD superconductor $T_{d}$-MoTe$_{2}$, which is considered to be a type-II Weyl semimetal \cite{SunY,Hasan1,GuguchiaMoTe2}. We had found a linear scaling of superfluid density at $T$ = 0 and the transition temperature $T_{\rm c}$, similar to the relation found earlier in the cuprates \cite{Uemura1}  and other Fe-based superconductors \cite{Luetkens,Khasanov2008,Carlo}.\\

Here, we report on the substantial increase of the superfluid density $n_{s}/m^{*}$ in van der Waals system 2H-NbSe$_2$ (\ref{fig1}(a)) under hydrostatic pressure, which seems not to be correlate with the suppression of CDW ordered state. We find the superfluid density in 2H-NbSe$_{2}$ to scale linearly with $T_{c}$. These results in combination to our previous results on MoTe$_{2}$ \cite{GuguchiaMoTe2} indicate that this linear relation has general validity for transition metal dichalcogenide superconductors. This relation is considered to be a hallmark feature of unconventional superconductivity \cite{Uemura1,Uemura3,EmeryKivelson} in cuprate and iron-pnictides superconductors. Our findings, therefore, pose a challenge for understanding the underlying quantum physics in these layered TMDs and might lead to a better understanding of generic aspects of non-BCS behaviors in unconventional superconductors.

In figure \ref{fig1}(b), we show the schematic phase diagram of the pressure dependence of the CDW transition temperature $T_{\rm CDW}$ and the superconducting transition temperature $T_{\rm c}$ for NbSe$_2$, according to reference \cite{QiCava}). The red arrows mark the pressures at which the $T$-dependence of the penetration depth and the superfluid density was measured. As it can be seen, the CDW transition is strongly reduced within the here investigated pressure range between $p$ = 0 to 2.2 GPa. In figures \ref{fig1}(c) and (d), two representative the transverse-field (TF) ${\mu}$SR-time spectra for 2H-NbSe$_{2}$ at ambient pressure and at the maximaum applied pressure of $p$ = 2.2 GPa are shown. Both measurements were performed in the pressure cell, with the constant background distracted. For both pressures, the transverse-field (TF) muon time spectra in a field of $\mu_0H =$ 70 mT are shown above ($T$ = 10 K) and below ($T$ = 0.25 K) the superconducting transition temperature $T_{{\rm c}}$. Above $T_{{\rm c}}$ the oscillations show a small relaxation due to the random local fields from the nuclear magnetic moments. Below $T_{{\rm c}}$ the relaxation rate strongly increases with decreasing temperature due to the presence of a nonuniform local magnetic field distribution as a result of the formation of a flux-line lattice (FLL) in the Shubnikov phase.

\begin{figure}[t!]
\centering
\includegraphics[width=1.4\linewidth]{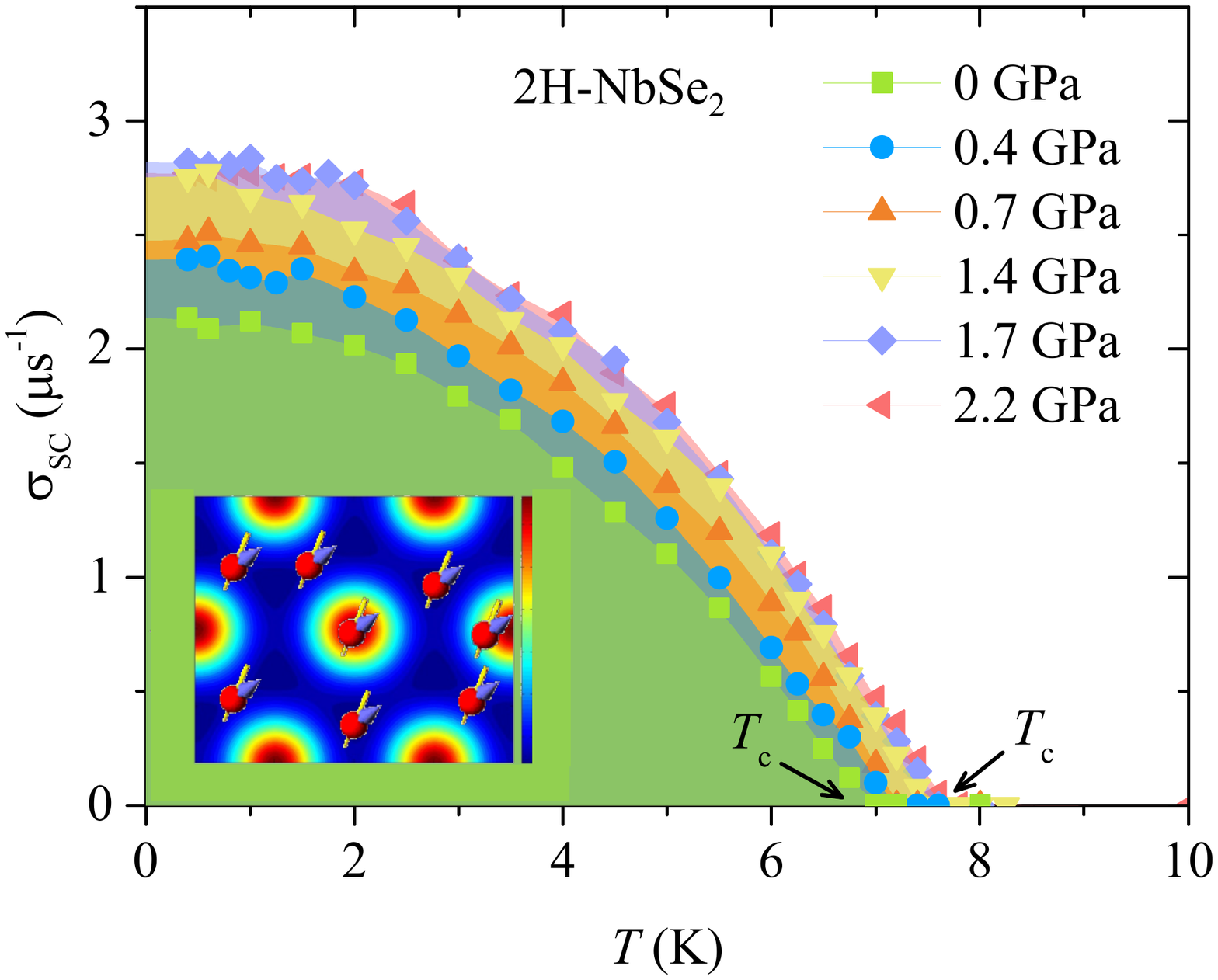}
\vspace{-1.7cm}
\caption{\textbf{Superconducting muon spin depolarization rate ${\sigma}_{\rm sc}$ for NbSe$_{2}$.} 
Temperature dependence of ${\sigma}_{\rm sc}$($T$), measured at various hydrostatic pressures in an applied magnetic field of ${\mu}_{\rm 0}H = 70$~mT.} 
\label{fig2}
\end{figure}

From the obtained TF muon time spectra, we have derived the temperature dependence of the muon spin depolarization rate ${\sigma}_{{\rm sc}}$, which is proportional to the second moment of the field distribution (see Method section). In figure \ref{fig2}, we show the temperature dependence of the muon spin depolarization rate ${\sigma}_{{\rm sc}}$ of NbSe$_{2}$ in the superconducting state at pressures of $p =$ 0, 0.4, 0.7, 1.4, and 2.2 GPa. Below $T_{{\rm c}}$ the relaxation rate ${\sigma}_{{\rm sc}}$ starts to increase from zero with decreasing temperature due to the formation of the FLL. We find the increase of the superconducting transition temperature $T_{{\rm c}}$ to be in good agreement with earlier reports \cite{YFeng,MLeroux}. Furthermore, we observed a substantial increase of the low-temperature value of the muon spin depolarization rate ${\sigma}_{{\rm sc}}$ with increasing pressure. This can be seen most clearly at base temperature, where the muon spin depolarization rate ${\sigma}_{{\rm sc}}$($T$ = 0.25 K) increases by almost ${\sim}$ 30 ${\%}$ from $p$ = 0 GPa to p = 2.2 GPa. Interestingly, at all pressures the form of the temperature dependence of ${\sigma}_{{\rm sc}}$, which reflects the topology of the SC gap, shows the saturation at low temperatures, indicating the presence of the isotropic pairing in NbSe$_{2}$ at all applied pressure. 
%


\begin{figure*}[t!]
\centering
\includegraphics[width=1.0\linewidth]{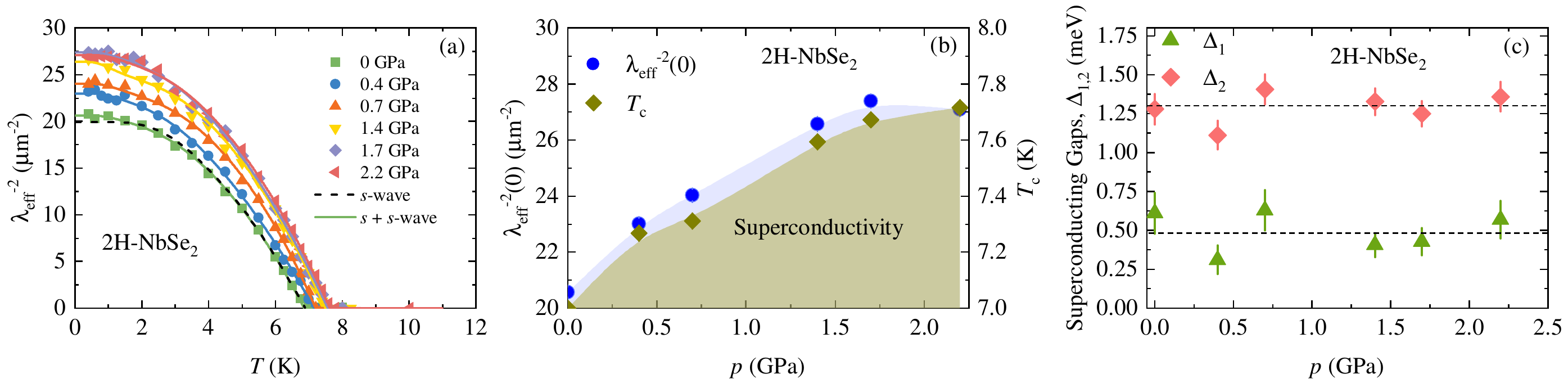}
\vspace{-0.5cm}
\caption{\textbf{Pressure evolution of various quantities.}
(a) The temperature dependence of ${\lambda}^{-2}$ measured at various applied hydrostatic pressures for NbSe$_{2}$. The solid lines correspond to a two-gap ($s+s$)-wave model, the dashed line represent a fit using a single gap $s$-wave model. (b) Pressure dependence of $T_{c}$ and the zero-temperature value of ${\lambda}^{-2}(0)$. (c) Pressure dependence of the zero-temperature values of the small superconducting gap ${\Delta}_1$ and the large superconducting gap ${\Delta}_2$.}
\label{fig3}
\end{figure*}

The second moment of the resulting inhomogeneous field distribution is related to the magnetic penetration depth $\lambda$ as $\left<\Delta B^2\right>\propto \sigma^2_{sc} \propto \lambda^{-4}$, whereas $\sigma_{sc}$ is the Gaussian relaxation rate due to the formation of  FLL \cite{Brandt}. In order to investigate the symmetry of the superconducting gap, we have therefore derived the temperature-dependent London magnetic penetration depth ${\lambda}(T)$, which is related to the relaxation rate by: 
\begin{equation}
\frac{\sigma_{sc}(T)}{\gamma_{\mu}}=0.06091\frac{\Phi_{0}}{\lambda^{2}(T)},
\end{equation}
Here, ${\gamma_{\mu}}$ is the gyromagnetic ratio of the muon, and ${\Phi}_{{\rm 0}}$ is the magnetic-flux quantum. Thus, the flat $T$-dependence of ${\sigma}_{{\rm sc}}$ observed at various pressures for low temperatures (see Fig.~\ref{fig2}) is consistent with a nodeless superconductor, in which
$\lambda^{-2}\left(T\right)$ reaches its zero-temperature value exponentially.

To proceed with a quantitative analysis, we consider the local (London) approximation (${\lambda}$ ${\gg}$ ${\xi}$, where ${\xi}$ is the coherence length) and employ the empirical ${\alpha}$-model. The model, widely used in previous investigations of the penetration depth of multi-band superconductors \cite{Bastian,Tinkham,carrington,padamsee,GuguchiaPRB}, assumes that the gaps occuring in different bands, besides a common $T_{{\rm c}}$, are independent of each other. The superfluid density is calculated for each component separately \cite{GuguchiaPRB} and added together with a weighting factor. For our purposes, a two-band model suffices, yielding: 
\begin{equation}
\frac{\lambda^{-2}(T)}{\lambda^{-2}(0)}=\omega_{1}\frac{\lambda^{-2}(T,\Delta_{0,1})}{\lambda^{-2}(0,\Delta_{0,1})}+\omega_{2}\frac{\lambda^{-2}(T,\Delta_{0,2})}{\lambda^{-2}(0,\Delta_{0,2})},
\end{equation}
Here ${\lambda}(0)$ is the London magnetic penetration depth at zero temperature, ${\Delta_{0,i}}$ is the value of the $i$-th SC gap ($i=1$, 2) at
$T=0$~K, and ${\omega}_{i}$ is the weighting factor, which measures their relative contributions to ${\lambda^{-2}}$ (i.e. ${\omega}_{1}+{\omega}_{2}=1$). \\
The results of this analysis are presented in Fig. \ref{fig3}a, where the temperature dependence of ${\lambda^{-2}}$ for NbSe$_{2}$ is plotted at pressures of $p =$ 0, 0.4, 0.7, 1.4, and 2.2 GPa. The dashed and the solid lines for ambient pressure results represent fits for the temperature-dependent London magnetic penetration at ambient pressure using a $s$-wave and a $s$ + $s$-wave model, respectively \cite{GuguchiaMoTe2}. As it can be seen, the $s$ + $s$-wave provides a much better description of the data, thereby ruling out a out the simple $s$-wave model as an adequate description of ${\lambda^{-2}}$($T$) for 2H-NbSe$_{2}$. The two gap $s$ + $s$-wave scenario with a small gap ${\Delta}_{1}$ ${\simeq}$ 0.55(3) meV  and a large gap ${\Delta}_{2}$ ${\simeq}$ 1.25(5) meV for $p$ = 0 GPa (with the pressure independent weighting factor of ${\omega}_{2}$ = 0.8(1)), describes the experimental data remarkably well. The presence of two isotopic gaps and their values are in very good agreement with previous results \cite{Le,Noat,Yokoya,Dvir,Borisenko}. According to ARPES data several independent electronic bands (four Nb-derived bands with roughly cylindrical Fermi surfaces, centered at the ${\Gamma}$ and K points and one Se-derived bands with a small ellipsoid pocket around the ${\Gamma}$ point) cross the Fermi surface in NbSe$_{2}$, two-gap superconductivity can be understood by assuming that the SC gaps open at two distinct types of bands. We find that two gap $s$ + $s$-wave superconductivity is preserved up to the highest applied pressure of $p$ = 2.2 GPa. All the $s$ + $s$-wave fits for all pressures are shown in figure \ref{fig3}(c). Furthermore, the pressure dependence of all the parameters extracted from the data analysis within the ${\alpha}$ model are plotted in Figs.~\ref{fig3}(b,c). The critical temperature $T_{\rm c}$ increases with pressure only by ${\sim}$ 0.7 K at the maximum applied pressure of $p$ = 2.2 GPa, as shown in Fig.~\ref{fig3}b. We, however, observe a substantial increase of the superfluid density ${\lambda^{-2}}$ with increasing pressures, as shown in figure \ref{fig3}(b). At the maximum applied pressure of $p$ = 2.2 GPa the increase of ${\lambda^{-2}}$ is $\Delta p \approx$ 31.4(8) ${\%}$ compared to the value at ambient pressure. The absolute size of the small gap ${\Delta}_{1}$ ${\simeq}$ 0.5(3) meV and the large gap ${\Delta}_{2}$ ${\simeq}$ 1.25(5) meV remain nearly unchanged by pressure, as shown in figure \ref{fig3}(c). \\

%

%
The London magnetic penetration depth ${\lambda}$ is given as a function of $n_{\rm s}$, $m^{*}$, ${\xi}$ and the mean free path $l$, according to \\
\begin{equation}
\begin{aligned}
\frac{1}{\lambda^2} = \frac{4\pi n_se^2}{m^*c^2} \times  \frac{1}{1 + \xi/l}, 
\end{aligned}
\end{equation} 
For systems close to the clean limit, ${\xi}$/$l$ ${\rightarrow}$ 0, the second term essentially becomes unity, and the simple relation 1/${\lambda}^{2}$ $\propto$ $n_{s}/m^{*}$ holds. Considering the upper critical fields $H_{c2}$ of 2H-NbSe$_{2}$, as reported in detail by Soto \textit{et al.} \cite{Soto}, we can estimate the in-plane coherence length to be ${\xi}_{ab}$ ${\simeq}$ 7.9 nm at ambient pressure $p$ = 0 GPa. At ambient pressure, the in-plane mean free path $l$ was estimated to be $l_{ab}$ ${\simeq}$ 183 nm \cite{Soto}. No estimates are currently available for $l$ under pressure. However, the in-plane $l$ is most probably independent of pressure, considering the fact that the effect of compression is mostly interlayered, i.e., the intralayer Nb-Se bond length remains nearly unchanged (especially in the here investigated pressure region).\cite{YFeng} This very small effect of compression can be attributed to the unique anisotropy resulting from the stacking of layers with van der Waals type interactions between them. Thus, in view of the short coherence length and relatively large $l$, we can reliably assume that NbSe$_{2}$ lies close to the clean limit \cite{GuguchiaMoTe2,Frandsen}. With this assumption, we obtain the ground-state value $n_{s}/(m^{*}/m_{e}$) ${\simeq}$ 5.7 ${\times}$ 10$^{27}$ m$^{-3}$, and 7.5 ${\times}$ 10$^{27}$ m$^{-3}$ for $p$ = 0 GPa, and 2.2 GPa respectively. 


\begin{figure}[b!]
\centering
\includegraphics[width=1.0\linewidth]{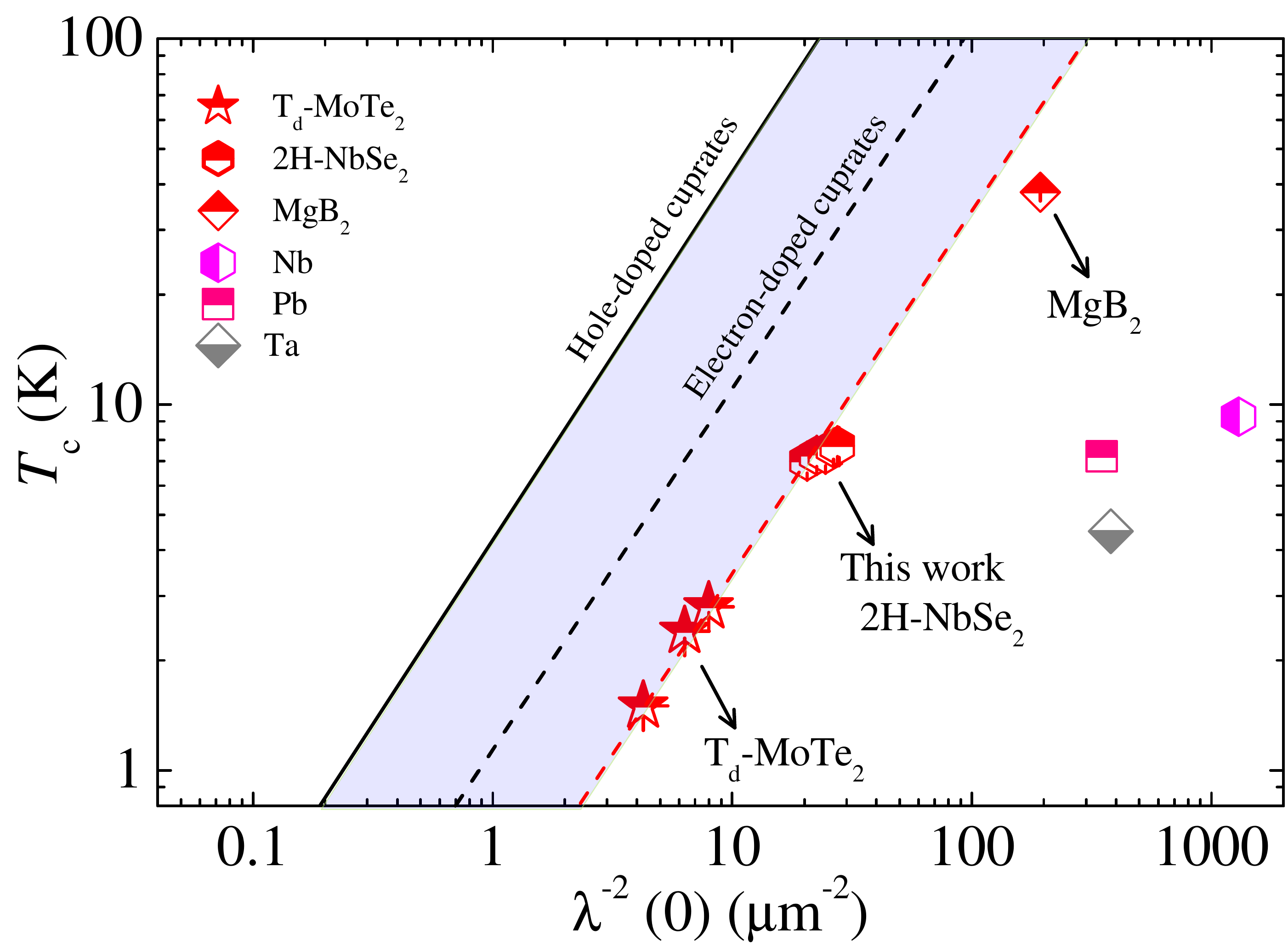}
\vspace{-0.5cm}
\caption{ \textbf{Superfluid Density versus $T_{\rm c}$.} 
A plot of $T_{\rm c}$ against the ${\lambda}^{-2}(0)$ obtained from our ${\mu}$SR experiments in 
NbSe$_{2}$ and MoTe$_{2}$. The dashed red line represents the linear fit to the MoTe$_{2}$ data. Uemura relation for hole and electron-doped cuprates are shown as solid \cite{Uemura1,Uemura3,Uemura4,Uemura5,Uemura6,Niedermayer} and dashed lines \cite{Shengelaya}, respectively. The points for various conventional BCS superconductors and for 2H-NbSe$_{2}$ are also shown.}
\label{fig4}
\end{figure}
%
The strong enhancement of the superfluid density ${\lambda}^{-2}(0)$ ${\propto}$ $n_{s}/(m^{*}/m_{e}$) in 2H-NbSe$_{2}$ under pressure, as discussed above, is an essential finding of this paper. It should be noted that the impairment of the charge-density wave ordering, and the associated lowering of the charge-density wave ordering transition $T_{CDW}$ under pressure may cause the restoring of some electronic density of states at the Fermi surface. However, the electron and hole states that condense into the CDW ordered state comprise only $\approx$ 1 ${\%}$ of the total density of states at the Fermi surface in 2H-NbSe$_2$ \cite{Straub,Rossnagel,Johannes}. The expected maximal increase of the total density of states caused by a complete suppression of the charge-density wave would be $\Delta D(E_{\rm F}) \approx$ 1 ${\%}$, which cannot solely attribute for the the observed ${\sim}$ 30 ${\%}$ enhancement of the superfluid density in 2H-NbSe$_2$. Thus, the large pressure effect on $n_{s}/(m^{*}/m_{e}$) has a more complex origin.
We also observed that both superconducting gaps ${\Delta}_{1}$ and ${\Delta}_{2}$ are nearly pressure independent up to a pressure of $p =$ 2.2 GPa (figure \ref{fig3}(c)), while the CDW transition temperature is largely reduced in this pressure range $p$ = 0 - 2.2 GPa \cite{YFeng,MLeroux}. This implies that the gap values are insensitive to the suppression of the CDW state. 
This observation strongly supports the idea that superconducting and CDW orders are somewhat isolated from each other and that CDW pairing have only minimal effect on superconductivity in 2H-NbSe$_{2}$ \cite{MLeroux}.
Furthermore, we find the superfluid density in 2H-NbSe$_{2}$ to scale linearly with $T_{c}$, as shown in figure \ref{fig4}, which is not expected within BCS theory. Remarkably, the ratio between the superfluid density and the critical temperature $T_{c}$ in 2H-NbSe$_{2}$ is nearly the same as for the TMD superconductor 1T'-MoTe$_{2}$ \cite{GuguchiaMoTe2}, indicating a common mechanism and related electronic origin for the superconductivity. \\

The nearly linear relationship between $T_{\rm c}$ and the superfluid density was originally observed in hole-doped cuprates \cite{Uemura1,Uemura3}, where the ratio between  $T_{\rm c}$ and their effective Fermi temperature $T_{\rm F}$ being about  $T_{\rm c}$/ $T_{\rm F}$ ${\sim}$ 0.05, which means about 4-5 times
reduction of $T_{\rm c}$ from the ideal Bose Condensation temperature for a non-interacting Bose gas. These results were discussed in terms of the crossover  from Bose Einstein Condensation to BCS-like condensation \cite{Uemura4,Uemura5,Uemura6}. Within the picture of BEC to BCS crossover, systems exhibiting small
$T_{\rm c}$/$T_{\rm F}$ (large $T_{\rm F}$) are considered to be in the BCS-like side, while the linear relationship between $T_{\rm c}$ and $T_{\rm F}$ is expected only in the BEC-like side. This relationship has in the past been used for the characterization of BCS-like, so-called conventional and BEC-like, so-called unconventional superconductors. The here present results on 2H-NbSe$_{2}$ together with our previously reported results on 1T'-MoTe$_{2}$ demonstrate that a linear relation between $T_{\rm c}$ and the superfluid density holds for these TMDs systems. However, we find the ratio $T_{\rm c}$/$T_{\rm F}$ to be reduced further by a factor of ${\sim}$ 20. These systems fall into the clean limit, therefore, the linear relation is unrelated to pair breaking, and can be regarded to hold between $T_{\rm c}$ and $n_{s}/m^{*}$. It should be noticed that the widely studied layered superconductor MgB$_{2}$, that has a layered structure, consisting of hexagonal boron layered, exhibits a ratio $T_{\rm c}$/$T_{\rm F}$ \cite{Niedermayer}, which is almost identical to the value found here for the TMDs. This implies that the BEC-like linear relationship may exist in systems with $T_{\rm c}$/$T_{\rm F}$ reduced even by a factor of 20 from the ratio in hole doped cuprates. \\

In refs \cite{Uemura5,Uemura6,Uemura7}, one of the present authors pointed out that there seem to exist at least two factors which determine $T_{\rm c}$ in unconventional superconductors: one is the superfluid density and the other is the closeness to the competing state.  The second factor can be seen in the energy of the magnetic resonance mode, which represents the difference in free energy between the superconducting state and the competing magnetically ordered state. In the case of hole-doped cuprates, the competing state is characterized by antiferromagnetic order, but frustrated by the introduction of doped holes.  In the case of electron-doped cuprates, the competing state develops in antiferromagnetic network diluted by the doped carriers.  In the case of present TMDC systems, the competing state comes from charge density waves.  These systematic difference of competing states might be related to the three different ratios of $T_{\rm c}$/$T_{\rm F}$ seen in the three different families of superconductors shown in Fig. 4.

In conclusion, we provide the first microscopic investigation of the superconductivity under hydrostatic pressure in the layered superconductor 2H-NbSe$_{2}$. Specifically, the zero-temperature magnetic penetration depth ${\lambda}_{eff}\left(0\right)$ and the temperature dependence of ${\lambda_{eff}^{-2}}$ were studied in 2H-NbSe$_{2}$ by means of ${\mu}$SR experiments as a function of pressure up to $p {\simeq}$ 2.2 GPa. The superfluid densities at all pressures are best described by a two gap $s$ + $s$-wave scenario. We find that the application of pressure causes a substantial increase of the superfluid density $n_{s}/m^{*}$. 
Moreover, we show that the superfluid density in 2H-NbSe$_{2}$ scales linearly with $T_{c}$, this behavior was previously attributed to unconventional superconductors. Our current results, in combination with previous results on 1T'-MoTe$_{2}$ \cite{GuguchiaMoTe2} demonstrate that a linear relation holds for these transition metal dichalcogenide superconductors but with a ratio $T_{\rm c}$/$T_{\rm F}$, which is reduced by a factor of 20 from the ratio in hole doped cuprates. This implies that the superfluid density of TMDs is about an order of magnitude higher than that in the cuprates with respect to their critical temperatures $T_{c}$. We also find that the values of the superconducting gaps are insensitive to the suppression of the CDW ordered state, indicating that CDW pairing has only a minimal effect on the superconductivity in 2H-NbSe$_{2}$. These results hint towards a common mechanism and electronic origin for superconductivity in TMDs, which might have far reaching consequences for the future development of devices based on these materials.
\section{METHODS}

\textbf{Sample preparation}: Single-phase polycrystalline samples of 2H-NbSe$_2$ were prepared by means of high-temperature solid-state synthesis. Stoichiometric amounts of niobium powder (99.99 \%) and selenium shots (99.999 \%) were mixed and heated in a sealed quartz tube, under an inert atmosphere, at 750 $^\circ$C for 3 days. In order to ensure high homogeneity, the samples were ground, pressed to a pellet, and reheated in a sealed quartz tube under an inert atmosphere at 750 $^\circ$C for another 5 days. Eventually, the samples were quenched in water to ensure no selenium loss. \\ 

\textbf{Pressure cell}:  Pressures up to 2.2 GPa were generated in a double wall piston-cylinder type of cell made of CuBe material, especially designed to perform ${\mu}$SR experiments under pressure \cite{GuguchiaPressure,Andreica,MaisuradzePC,GuguchiaNature}. As a pressure transmitting medium Daphne oil was used. The pressure was measured by tracking the SC transition of a very small indium plate by AC susceptibility. The filling factor of the pressure cell was maximized. The fraction of the muons stopping in the sample was approximately 40 ${\%}$.\\

\textbf{${\mu}$SR experiment}: 

In a ${\mu}$SR experiment nearly 100 ${\%}$ spin-polarized muons ${\mu}$$^{+}$ are implanted into the sample one at a time \cite{Sonier}. The positively charged ${\mu}$$^{+}$ thermalize at interstitial lattice sites, where they act as magnetic microprobes. In a magnetic material the muon spin precesses in the local field $B_{\rm \mu}$ at the muon site with the Larmor frequency ${\nu}_{\rm \mu}$ = $\gamma_{\rm \mu}$/(2${\pi})$$B_{\rm \mu}$ (muon gyromagnetic ratio $\gamma_{\rm \mu}$/(2${\pi}$) = 135.5 MHz T$^{-1}$). Using the $\mu$SR technique important length scales of superconductors can be measured, namely the magnetic penetration depth $\lambda$ and the coherence length $\xi$. If a type II superconductor is cooled below $T_{\rm c}$ in an applied magnetic field ranged between the lower ($H_{c1}$) and the upper ($H_{c2}$) critical fields, a vortex lattice is formed which in general is incommensurate with the crystal lattice with vortex cores separated by much larger distances than those of the unit cell. Because the implanted muons stop at given crystallographic sites, they will randomly probe the field distribution of the vortex lattice. Such measurements need to be performed in a field applied perpendicular to the initial muon spin polarization (so called TF configuration). 

${\mu}$SR experiments under pressure were performed at the ${\mu}$E1 beamline of the Paul Scherrer Institute (Villigen, Switzerland, where an intense high-energy ($p_{\mu}$ = 100 MeV/c) beam of muons is implanted in the sample through the pressure cell. The low background GPS (${\pi}$M3 beamline) and low-temperature LTF instruments were used to study the single crystalline as well as the polycrystalline samples of MoTe$_{2}$ at ambient pressure.\\
  
\textbf{Analysis of TF-${\mu}$SR data}: 

 The TF ${\mu}$SR data were analyzed by using the following functional form:\cite{Bastian}
\begin{equation}
\begin{aligned}
P(t)=A_s\exp\Big[-\frac{(\sigma_{sc}^2+\sigma_{nm}^2)t^2}{2}\Big]\cos(\gamma_{\mu}B_{int,s}t+\varphi) \\
 + A_{pc}\exp\Big[-\frac{\sigma_{pc}^2t^2}{2}\Big]\cos(\gamma_{\mu}B_{int,pc}t+\varphi), 
\end{aligned}
\end{equation}
 Here $A_{\rm s}$ and $A_{\rm pc}$  denote the initial assymmetries of the sample and the pressure cell, respectively. $\gamma/(2{\pi})\simeq 135.5$~MHz/T 
is the muon gyromagnetic ratio, ${\varphi}$ is the initial phase of the muon-spin ensemble and $B_{\rm int}$ represents the internal magnetic field at the muon site. The relaxation rates ${\sigma}_{\rm sc}$ and ${\sigma}_{\rm nm}$ characterize the damping due to the formation of the FLL in the SC state and of the nuclear magnetic dipolar contribution, respectively. In the analysis ${\sigma}_{\rm nm}$ was assumed to be constant over the entire temperature range and was fixed to the value obtained above $T_{\rm c}$ where only nuclear magnetic moments contribute to the muon depolarization rate ${\sigma}$. The Gaussian relaxation rate, ${\sigma}_{\rm pc}$, reflects the depolarization due to the nuclear moments of the pressure cell. The width of the pressure cell signal increases below $T_{c}$. As shown previously \cite{MaisuradzePC}, this is due to the influence of the diamagnetic moment of the SC sample on the pressure cell, leading to the temperature dependent ${\sigma}_{\rm pc}$ below $T_{c}$. In order to consider this influence we assume the linear coupling between ${\sigma}_{\rm pc}$ and the field shift of the internal magnetic field in the SC state:\\
${\sigma}_{\rm pc}$($T$) = ${\sigma}_{\rm pc}$($T$ ${\textgreater}$ $T_{\rm c}$) + $C(T)$(${\mu}_{\rm 0}$$H_{\rm int,NS}$ - ${\mu}_{\rm 0}$$H_{\rm int,SC}$), where  ${\sigma}_{\rm pc}$($T$ ${\textgreater}$ $T_{\rm c}$) = 0.25 ${\mu}$$s^{-1}$ is the temperature independent Gaussian relaxation rate. ${\mu}_{\rm 0}$$H_{\rm int,NS}$ and ${\mu}_{\rm 0}$$H_{\rm int,SC}$ are the internal magnetic fields measured in the normal and in the SC state, respectively. 
As indicated by the solid lines in Figs.~2b,c the ${\mu}$SR data are well described by Eq.~(1). The good agreement between the fits and the data demonstrates that the model used describes the data rather well.\\

\textbf{Analysis of ${\lambda}(T)$}:



${\lambda}$($T$) was calculated within the local (London) approximation (${\lambda}$ ${\gg}$ ${\xi}$) by the following expression \cite{Bastian,Tinkham}:
\begin{equation}
\frac{\lambda^{-2}(T,\Delta_{0,i})}{\lambda^{-2}(0,\Delta_{0,i})}=
1+\frac{1}{\pi}\int_{0}^{2\pi}\int_{\Delta(_{T,\varphi})}^{\infty}(\frac{\partial f}{\partial E})\frac{EdEd\varphi}{\sqrt{E^2-\Delta_i(T,\varphi)^2}},
\end{equation}
where $f=[1+\exp(E/k_{\rm B}T)]^{-1}$ is the Fermi function, ${\varphi}$ is the angle along the Fermi surface, and ${\Delta}_{i}(T,{\varphi})={\Delta}_{0,i}{\Gamma}(T/T_{\rm c})g({\varphi}$)
(${\Delta}_{0,i}$ is the maximum gap value at $T=0$). The temperature dependence of the gap is approximated by the expression ${\Gamma}(T/T_{\rm c})=\tanh{\{}1.82[1.018(T_{\rm c}/T-1)]^{0.51}{\}}$,\cite{carrington} while $g({\varphi}$) describes the angular dependence of the gap and it is replaced by 1 for both an $s$-wave and an $s$+$s$-wave gap, and ${\mid}\cos(2{\varphi}){\mid}$ for a $d$-wave gap.\\
%
\section{Acknowledgments}~
The ${\mu}$SR experiments were carried out at the Swiss Muon Source (S${\mu}$S) Paul Scherrer Insitute, Villigen, Switzerland. Z. Guguchia gratefully acknowledges the financial support by the Swiss National Science Foundation (SNFfellowship P300P2-177832). The work at the University of Zurich was supported by the Swiss National Science Foundation under Grant No. PZ00P2\_174015. Work at Department of Physics of Columbia University is supported by US NSF DMR-1436095 (DMREF) and NSF DMR-1610633. M.Z.H. is supported by US DOE/BES grant no. DE-FG-02-05ER46200. R.K. acknowledges the Swiss National Science Foundation (grants 200021\_149486 and 200021\_175935). A.N. acknowledges funding from the European Union's Horizon 2020 research and innovation programme under the Marie Sklodowska-Curie grant agreement No 701647.

\end{document}